\documentclass[aip,jcp,10pt,floatfix,twocolumn]{revtex4-2} 
\usepackage{color,soul} 
\usepackage{graphicx}
\usepackage{amsmath, amsthm, amssymb, mathtools}
\usepackage{multirow} 
\usepackage[normalem]{ulem}

\usepackage{dcolumn}% Align table columns on decimal point
\usepackage{bm}% bold math
\usepackage[utf8]{inputenc}
\usepackage[T1]{fontenc}
\usepackage{mathptmx}
\usepackage{etoolbox}

\DeclareMathOperator{\Tr}{Tr}
\usepackage{braket}

\usepackage[most]{tcolorbox} % The `most' library loads commonly used libraries
\usepackage{xcolor} % Optional: if you want more flexibility with colors

\usepackage{hyperref}
\hypersetup{
colorlinks=true,
linkcolor=blue,
citecolor=black,
urlcolor=black
}

\newtcolorbox{mybox}[1][]{%
colback=gray!10!white, % Light gray background
colframe=gray!50!black, % Gray frame
fonttitle=\bfseries,
title=#1,
sharp corners,
boxrule=0.5mm,
boxsep=5pt,
breakable
}

\usepackage{xcolor} % Ensure this is in your preamble

\newif\ifhighlight
\highlightfalse
\DeclareRobustCommand{\highlighttext}[1]{%
  \ifhighlight
    \textcolor{red}{#1}%
  \else
    #1%
  \fi
}

\makeatletter
\def\@email#1#2{%
\endgroup
\patchcmd{\titleblock@produce}
{\frontmatter@RRAPformat}
{\frontmatter@RRAPformat{\produce@RRAP{*#1\href{mailto:#2}{#2}}}\frontmatter@RRAPformat}
{}{}
}%
\makeatother
\begin{document}

%\preprint{AIP/123-QED}

\title{Ultrastrong coupling limit to quantum mean force Gibbs state for anharmonic environment}

\author{Prem Kumar}
\email{premkr@imsc.res.in}
\affiliation{Optics and Quantum Information Group, The Institute of Mathematical Sciences, C.I.T. Campus, Taramani, Chennai 600113, India.}
\affiliation{Homi Bhabha National Institute, Training School Complex, Anushakti Nagar, Mumbai 400094, India}

\author{Sibasish Ghosh}
%\email{sibasish@imsc.res.in}
\affiliation{Optics and Quantum Information Group, The Institute of Mathematical Sciences, C.I.T. Campus, Taramani, Chennai 600113, India.}
\affiliation{Homi Bhabha National Institute, Training School Complex, Anushakti Nagar, Mumbai 400094, India}

\date{\today}

\begin{abstract}
The equilibrium state of a quantum system can deviate from the Gibbs state if the system-environment (SE) coupling is not weak. An analytical expression for this mean force Gibbs state (MFGS) is known in the ultrastrong coupling (USC) regime for the Caldeira-Leggett (CL) model that assumes a harmonic environment. Here, we derive analytical expressions for the MFGS in the USC regime for more general SE models. For all the generalized models considered here, we find the USC state to be diagonal in the basis set by the SE interaction, just like in the CL case. While for the generic model considered, the corresponding USC-MFGS is found to alter from the CL-result, we do identify a class of models more general than the CL model for which the CL-USC result remains unchanged. We also provide numerical verification for our results. These results provide key tools for the study of strong coupling quantum thermodynamics and several quantum chemistry and biology problems under more realistic SE models, going beyond the CL model.
\end{abstract}
\maketitle

\section{Introduction}
Non-negligible system-environment (SE) coupling can cause the equilibrium state of a system to deviate from the textbook Gibbs state.\cite{Miller2018, kirkwood1935statistical} The concept of Mean Force Gibbs State (MFGS) has been developed to capture the idea of this generalized equilibrium state, and has historically seen a lot of application in the field of chemistry \cite{allen2006molecular, maksimiak2003molecular, roux1999implicit, roux1995calculation} and, more recently, to the theory of strong coupling quantum thermodynamics.\cite{Miller2018, seifert2016first, philbin2016thermal, jarzynski2004nonequilibrium, jarzynski2004nonequilibrium, campisi2009fluctuation, rivas2020strong, talkner2020colloquium, strasberg2020measurability} Many of these approaches to calculate the quantum MFGS assume the Caldeira-Leggett (CL) model for the system and the environment, in which the environment is modelled as a set of harmonic oscillators coupled to the system.

The harmonic environment approximation has been widely successful in capturing the behavior of a variety of physical systems.\cite{feynman2000theory, Leggett1987dynamics, gilmore2006criteria, nazir2009correlation, Xu_1994, Ishizaki_2009, Rebentrost_2009, de_Vega_2011, Mohseni_2013, Tang_1993, Barbara_1996, Mavros_2016, Zhao_2012, Tanimura_1993, weiss2012quantum, BRE02, gardiner2004quantum, bose2023impact} Yet, advancement in experimental techniques and more detailed theoretical considerations have shown the inadequacy of the harmonic environment approximation, sparking  numerous studies on open quantum system with anharmonic environment.\cite{okumura1996unified, evans2000anharmonic, wang2004semiclassical, lockwood2002effects, Wang_2019, Ilk_1994, Shi_2004, ohmine1993fluctuation, thoss2006quantum, wang2007quantum, bramberger2020dephasing, rognoni2021caldeira, gottwald2015applicability} For example, for electron transfer happening in the presence of strongly coupled low-frequency intramolecular modes or an environment consisting of nonpolar liquids, the harmonic approximation is known to fail.\cite{evans2000anharmonic, lockwood2002effects, wang2004semiclassical,wang2007quantum, thoss2006quantum, ohmine1993fluctuation, okumura1996unified, bramberger2020dephasing}
It is hence worth investigating the MFGS result for SE models beyond the CL model.

If $\hat{H}_{\text{SE}}$ denotes the full Hamiltonian of a system and environment, then the MFGS is formally defined as \cite{cresser2021weak, trushechkin2022open}
\begin{align}\label{eqn:definition of MFGS}
\hat{\rho} = Z^{-1}\Tr_{E}\left[ e^{-\beta \hat{H}_{\text{SE}}}\right],
\end{align}
where $Z$ is an appropriate normalization constant.

For the CL model, analytical expression for the MFGS in ultrastrong coupling (USC) regime has been determined for continuous variable (CV) systems \cite{ankerhold2003phase, hilt2011hamiltonian} and, more recently, for a more general CL model (consisting of an arbitrary system Hamiltonian and coupling-operator).\cite{cresser2021weak} Moreover, this state is also confirmed to be the steady state of an ultrastrong coupling master equation.\cite{trushechkin2022quantum}

The USC-MFGS acts as a good indicator for the deviation of the MFGS from the textbook Gibbs state at large coupling. However, since this result is only valid for the CL model, its applicability is significantly limited. It also remains uncertain as to which aspects of the USC result are artifacts of the specific structures of the CL model, and which aspects might persist when the model is generalized. Here, we attempt to address this issue by studying the USC-MFGS for models beyond the CL model.

We find that for all the SE models considered by us, one of the core features of the CL-USC result, that the MFGS is diagonal in the basis set by the SE interaction, remains intact, although the functional form of the MFGS does get affected. For example, the most general extension of the CL model considered by us, which we call the GCL model, shows such deviations from the CL-USC result. Within the GCL model, we identify a subclass of models, which we call the GCL2 model, for which the CL-USC result remains exactly valid. Finally, we derive an analytical expression for the USC-MFGS for another class of model (distinct from the GCL model), the so-called Zwanzig model \cite{zwanzig1973nonlinear, Banerjee_2002, Banerjee_2002_Kramers, Barik_2003}, and find that the corresponding USC state, although still diagonal in the basis set by the SE interaction, has qualitative distinctions from the GCL2-USC result.

In Section~\ref{sec:USC-MFGS in CL model}, we rederive the known USC-MFGS result for the CL model \cite{cresser2021weak} using the path integral approach, to illustrate the method that we will be using to generalize the same result. Section~\ref{sec:generalization 1} contains the derivation of the USC result for the GCL model, while Section~\ref{sec:GCL2} acts as a special case of the same derivation for the more restricted GCL2 model. Finally, in Section~\ref{sec:beyond CL model}, we provide the USC-MFGS derivation for the Zwanzig model. Section~\ref{sec:Numerical verification} provides some numerical verifications of the analytical results derived here, and Section~\ref{sec:Conclusion} provides conclusions and future directions.

\section{USC-MFGS in CL model} \label{sec:USC-MFGS in CL model}
Consider a typical CL model in which a generic quantum system, with free Hamiltonian $\hat{H}_S$, has a one-to-one interaction with a set of harmonic oscillators that are uncoupled from each other. The full SE Hamiltonian is hence given by
\begin{align}
\hat{H}_{\mathrm{SE}} &= \hat{H}_S + \hat{H}_I, \label{eqn:discrete hamiltonian partition}\\
\hat{H}_I &= \sum_k\left[\frac{\hat{p}_k^2}{2 m_k}+\frac{1}{2} m_k \omega_k^2\left(\hat{q}_k- \alpha_k \hat{A}\right)^2\right],\label{eqn:discrete CL_hamiltonia}
\end{align}
where $\hat{A}$ is a generic system operator through which the system couples to the environment (which from now onwards we will refer to as `system coupling-operator') and
\begin{align}
\alpha_k \equiv \frac{c_k}{m_k \omega_k^2}, \label{eqn:definition of alpha_k} %eqn 3.379 BR02
\end{align}
where $m_k \text{ and } \omega_k$ are the mass and frequency of the $k$th harmonic oscillator and $c_k$ is the corresponding SE coupling strength. For a CV system, typically we have $\hat{A} = \hat{q}$, where $\hat{q}$ is the position operator of the system; while for the spin-Boson model, $\hat{A}$ is taken to be a Pauli matrix.\cite{weiss2012quantum, BRE02, gardiner2004quantum}

The MFGS (Eq.~\ref{eqn:definition of MFGS}) can be calculated using the Feynman path integral \cite{feynman1965path} in the basis of the operator $\hat{A}$ by evolving the system and the environment in the `imaginary time' by an amount $\beta$ and then tracing over the environment degrees of freedom.\cite{feynman2000theory, weiss2012quantum}

Note a subtlety that arises when the system coupling-operator $\hat{A}$ has degeneracy, and hence each of its eigenstates is not uniquely labelled by an eigenvalue. To fix this problem, let us introduce an operator $\hat{\phi}$ that lifts the degeneracy in $\hat{A}$. That is, $[\hat{\phi}, \hat{A}] = 0$ and $\hat{A} + \hat{\phi}$ has no degeneracy. The final result will not depend upon the particular choice of $\hat{\phi}$. Without any loss of generality and for convenience, we will also assume that $\hat{\phi}$ is itself non-degenerate, so that its eigenvalues, $(a,b)$, can be used as a unique label to express the system's reduced density matrix elements as $\hat{\rho}(a,b)$.

Let a generic imaginary time path be denoted as $\{A(t),\phi(t), q_k(t)\}$, where $A(t) \in \{A_i\}, $
$\phi(t) \in \{\phi_i\}, $
$\text{and } q_k(t) \in \{{q_k}_i\}$ 
, where $\{A_i\},$ 
$\{\phi_i\}$ 
and $\{{q_k}_i\}$ 
are the sets of eigenvalues of operators $\hat{A},$
$\hat{\phi}$ 
and $\hat{q}_k$, respectively.
We emphasize here that $A(t)$ denotes a path that the system can take in the path integral and should not be confused to mean that the system coupling-operator, $\hat{A}$, has a time dependence. In fact, $\hat{A}$ has been assumed to be time-independent throughout this paper.

Now, let $S_S[A(t), \phi(t)]$ and $S_I[A(t),q_k(t)]$ represent the imaginary time action corresponding to $\hat{H}_S$ and $\hat{H}_I$, respectively. Note how $S_I$ is not a function of $\phi(t)$ while $S_S$ is. This is because while $\hat{H}_S$ is an arbitrary system operator, $\hat{H}_I$ has system dependence only through the operator $\hat{A}$, and hence the corresponding action, $S_I$, cannot distinguish within the degenerate subspace of $\hat{A}$, which removes the $\phi(t)$ dependence.

Also note that the paths $A(t)$ and $\phi(t)$ are not independent of each other. That is, $A(t)$ and $\phi(t)$ denote some eigenvalues of the operators $\hat{A}$ and $\hat{\phi}$, respectively, such that one can find a common eigenvector $\ket{v(t)} \equiv \ket{A(t), \phi(t)}$ such that $\hat{A} \ket{v(t)} = A(t) \ket{v(t)}$ and $\hat{\phi}\ket{v(t)} = \phi(t) \ket{v(t)}$. But we can always treat $A(t)$ and $\phi(t)$ as independent of each other if we assume that the action $S_S[A(t), \phi(t)]$ is constructed in such a manner that, given some $A(t)$, it rules out all unphysical possibilities for the path $\phi(t)$ by assigning the corresponding action an infinite weight. For all the arguments in this paper, we \textit{don't} need to go into any further details on how to construct such a functional $S_S[A(t), \phi(t)]$.

The full SE path integral is given as
\begin{align}
\hat{\rho}(a, b, a_k, b_k) &= \Tilde{Z}^{-1} \left( \prod_k \int_{a_k}^{b_k} \mathcal{D}q_k(t) \right)\int_{\Tilde{a}}^{\Tilde{b}} \mathcal{D}A(t)  \nonumber \\
&\quad \times\int_{a}^{b} \mathcal{D}\phi(t) e^{- S_S[A(t), \phi(t)] - S_I[A(t),q_k(t)]} \label{eqn: CL full pathe integral}.
\end{align}
Here $\Tilde{Z}$ is the overall normalization factor and $a$, $b$, $a_k$, $b_k$, $\Tilde{a}$, and $\Tilde{b}$ correspond to $\phi(0)$, $\phi(\beta)$, $q_k(0)$, $q_k(\beta)$, $A(0)$, and $A(\beta)$, respectively. Tracing over the environment will now give us the MFGS as
\begin{align}
\hat{\rho}_{\text{USC}}(a, &b) = \left( \prod_k \int_{-\infty}^{\infty} d a_k \right) \hat{\rho}(a, b, a_k, a_k)\\
&= \Tilde{Z}^{-1} \int_{\Tilde{a}}^{\Tilde{b}} \mathcal{D}A(t) \int_{a}^{b} \mathcal{D}\phi(t) e^{ -S_S[A(t), \phi(t)]} I[A(t)] \label{eqn:discrete full path integral}.
\end{align}
Here $I[A(t)] \equiv \prod_k I_k[A(t)]$ is called the influence functional \cite{feynman2000theory, weiss2012quantum},
where $I_k[A(t)]$ encapsulates the effect of the $k$th environment particle only and is defined as
\begin{align}
I_k[A(t)] &= \int_{-\infty}^{\infty} dx \int_{x}^{x} \mathcal{D}q_k(t) e^{- S_k\left[q_k(t), A(t)\right]}\label{eqn:environment side path integral}.
\end{align}
Here the action, $S_k$, is defined as
\begin{align}
S_k\left[q_k(t), A(t)\right] &\equiv \int_0^\beta dt \bigg\{ \frac{m_k}{2} \dot{q}_k(t)^2 \nonumber \\
&\quad + \frac{m_k\omega_k^2}{2} \left(q_k(t) - \alpha_k A(t)\right)^2\bigg\}\label{eqn:CL environment action}.
\end{align}
The explicit expression for $I[A(t)]$ for CL model \cite{feynman2000theory, weiss2012quantum} can be used to derive the CL-USC result.\cite{ankerhold2003phase, kumar2024local} But here, since we need to eventually generalize this result beyond the CL model, we will derive the expression for the USC-MFGS without explicitly evaluating $I[A(t)]$ for any specific model.

Let us choose an arbitrarily large but finite length scale $\Delta$ such that whenever $|q_k(t) - \alpha_k A(t)| > \Delta$ for any $t$, the potential energy (PE) cost of the corresponding path, $q_k(t)$, will be very high, and hence its contribution to the path integral will be negligible. We can hence write, for all paths $q_k^*(t)$ that \textit{do} have significant contribution to the path integral,
\begin{align}
| q_k^*(t) - \alpha_k A(t) | < \Delta \label{eqn:PE minimizing path}.
\end{align}
Let us define
\begin{align}
\braket{A} &\equiv \frac{ 1 }{\beta} \int_0^\beta dt A(t) \label{eqn:mean definition},\\
\braket{A^2} &\equiv \frac{ 1 }{\beta} \int_0^\beta dt A(t)^2 \label{eqn:squared mean definition},\\
\sigma_A^2 &\equiv \braket{A^2} - \braket{A}^2 \label{eqn: variance definition}.
\end{align}
Then, in Appendix.~\ref{Appendix: PE KE cost proof}, we have proven that if $\sigma_A^2 \gg \Delta^2/\alpha_k^2$, then
\begin{align}
\frac{1}{\beta}\int_0^\beta dt \dot{q}_k^*(t)^2 &\geq \alpha_k^2 \mu \frac{\sigma_A^2}{\beta^2} \label{eqn:kinetic energy cost}.
\end{align}
Here $\mu > 0$ is some constant. Now, note that in the USC limit, defined as
\begin{align}
\lim c_k \to \infty , \label{eqn:USC limit}
\end{align}
we have  $\alpha_k \to \infty$ (Eq.~\ref{eqn:definition of alpha_k}). Hence, in this limit, unless $\sigma_A = O(\Delta/\alpha_k)$, the RHS in Eq.~\ref{eqn:kinetic energy cost} diverges, causing the kinetic term $\dot{q}_k^*(t)^2$ in the action $S_k\left[q_k(t), A(t)\right]$ (Eq.~\ref{eqn:CL environment action}) to diverge. Hence, in the USC limit (Eq.~\ref{eqn:USC limit}), paths $q_k^*(t)$ (which have been defined to have convergent PE cost) will have divergent kinetic energy cost unless $\sigma_A$ vanishes, or in other words, unless we have
\begin{equation}
A(t) = A(0) = \Tilde{a}, \quad \text{for some constant} \; \Tilde{a} \label{eqn:frozen paths}.
\end{equation}

\begin{figure}[htbp]
\includegraphics[width=0.48\textwidth]{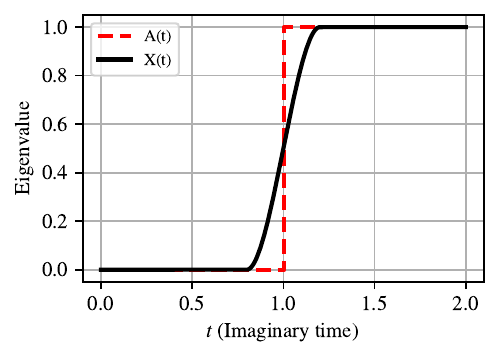}
\caption{ Approximating a piecewise constant function, $A(t)$, with a continuous and differentiable function $X(t)$. The function $X(t)$ is equal to $A(t)$ everywhere except near the discontinuous jump in $A(t)$ (when $1 - \delta< t <1 + \delta$, for $\delta = 0.2$). Near $t=1$, $X(t)$ takes the form of a cubic function, $f(t') = a_0 + a_1 t' + a_3 t'^3$, with $t' = t- 1$, $a_0 = 1/2 $, $a_1 = 3/ (4 \delta)$ and $a_3 = -1/(4 \delta^3)$. This means that we have $\lim_{\delta \to 0} X(t) = A(t)$. }
\label{fig:At_and_Xt_comparison}
\end{figure}

As we will see, Eq.~\ref{eqn:frozen paths} will greatly simplify the evaluation of the MFGS path integral (Eq.~\ref{eqn:discrete full path integral}). But before getting into that, we need to address a subtlety. If the system lives in a discrete Hilbert space, then $A(t)$ may not be continuous and differentiable everywhere, leading to potential mathematical irregularities. Although the results and proofs in this paper only require the paths \(A(t)\) to be square integrable (and not continuous and differentiable), we note that these mathematical irregularities can be addressed by substituting \(A(t)\) with a continuous and differentiable function \(X(t)\) that closely approximates it, such that the induced error, $|I[A(t)] - I[X(t)]|$, is arbitrarily small. For instance, Fig.~\ref{fig:At_and_Xt_comparison} illustrates how $X(t)$, using a cubic polynomial near $t=1$, smoothly approximates $A(t)$'s discrete jump.

Now, given Eq.~\ref{eqn:frozen paths}, the expression for the influence functional (Eq.~\ref{eqn:environment side path integral}) simplifies as
\begin{align}
I_k[\Tilde{a}] &= \int_{-\infty}^{\infty} dx \int_{x}^{x} \mathcal{D}q_k(t) e^{-S_k\left[q_k(t), \Tilde{a}\right]}\\
&= \int_{-\infty}^{\infty} dx \int_{x-\alpha_k \Tilde{a}}^{x- \alpha_k \Tilde{a}} \mathcal{D}x_k(t) e^{-S_k\left[x_k(t), 0 \right]} \label{eqn:cl case I_k calc step 2}\\
&= \int_{-\infty}^{\infty} dx \int_{x}^{x} \mathcal{D}x_k(t) e^{-S_k\left[x_k(t), 0 \right]}\\
&= I_k, \quad \text{which is independent of}\; \Tilde{a}.
\end{align}
Here in Eq.~\ref{eqn:cl case I_k calc step 2}, we defined $x_k(t) \equiv q_k(t) - \alpha_k \Tilde{a}$ and we have used the relation $S_k\left[q_k(t), \Tilde{a}\right] = S_k\left[x_k(t), 0 \right]$ (Eq.~\ref{eqn:CL environment action}). We hence have that $I[\Tilde{a}] \equiv I $ is independent of $\Tilde{a}$. Hence, the only contribution of the environment side path integral is to eliminate all the system side paths that do not satisfy Eq.~\ref{eqn:frozen paths}. This means that jumps within a degenerate subspace are still allowed and with equal weight as far as the environment side path integral is concerned. Therefore, the full MFGS path integral (Eq.~\ref{eqn:discrete full path integral}) simplifies as
\begin{align}
\hat{\rho}_{\text{USC}}(a,b) &= \Tilde{Z}^{-1} I \int_{a}^{b} \mathcal{D}\phi(t) \exp \left\{- S_S[\Tilde{a}, \phi(t)] \right\}\label{eqn:CL USC-MFGS integral form}.
\end{align}
Here $a \text{ and }b$ are the eigenvalues of the operator $\hat{\phi}$ and $\Tilde{a}$ is the eigenvalue of $\hat{A}$.\cite{abandtilda}
Then, the USC-MFGS is given as
\begin{align}
\hat{\rho}_{\text{USC}} &= Z^{-1} \exp \left\{ -\beta \Tilde{H} \right\} \label{eqn:general USC state},\\
\Tilde{H} &\equiv \sum_i \hat{P}_i \hat{H}_S \hat{P}_i. \label{eqn_CL_USC_reduced_H}
\end{align}
Here $\hat{P}_i$ is the projection operator on the $i$th eigensubspace of the system coupling-operator, $\hat{A}$, and $Z$ is the normalization constant. Eq.~\ref{eqn:general USC state} is equivalent to the USC-MFGS derived by Cresser and Anders \cite{cresser2021weak} using perturbation theory, and reduces to the USC-MFGS derived earlier for a CV system.\cite{ankerhold2003phase} Note that the derivation provided in this section is applicable for a generic CL model, discrete or continuous. See Appendix.~\ref{appendix_CL_USC_result_for_CV_system} for more details.

\section{USC-MFGS for Generalized Caldeira-Leggett model}\label{sec:generalization 1}
The generalization of the CL model has been considered by several authors \cite{Wang_2019, bramberger2020dephasing, Ilk_1994, wang2004semiclassical, wang2007quantum, Shi_2004} and can be characterized by an interaction Hamiltonian, $\hat{H}_I$ (Eq.~\ref{eqn:discrete CL_hamiltonia}), of the form
\begin{align}
\hat{H}_I &= \sum_k\left[\frac{\hat{p}_k^2}{2 m_k} + V_k(\hat{q}_k, c_k \hat{A}) \right]. \label{eqn:more general interaction hamiltonian}
\end{align}
Here $V_k(x,y)$ is a potential function bounded from below with a constraint that it should not renormalize the free effective system Hamiltonian $\hat{H}_S$. \cite{weiss2012quantum, wang2004semiclassical, bramberger2020dephasing, bhadra2016system} This is usually ensured by adding a so-called counter term in the potential, $V_k(x,y)$, which compensates for the shift in $\hat{H}_S$ because of the environment. Here, we assume that such a counter term has already been incorporated into $V_k(x,y)$, which translates into demanding that for a fixed value of $y$, if $x = \Tilde{x}_k(y)$ is the point at which $V_k(x,y)$ has a global minimum, then \cite{weiss2012quantum, bramberger2020dephasing}
\begin{equation}
V_k(\Tilde{x}_k(y),y) = 0 \label{eqn:GCL minima point}.
\end{equation}

These generalized CL models have been studied in various contexts. For instance, they have been used to explore the influence of an anharmonic environment on the non-Markovian dynamics of open systems \cite{bramberger2020dephasing}, to investigate certain condensed phase processes such as electron-transfer reactions \cite{Ilk_1994, Shi_2004, wang2007quantum, Wang_2019}, and to examine the electronic absorption spectra of chromophores embedded in condensed phase environments.\cite{wang2004semiclassical}

Here, we will generalize the derivation for the USC-MFGS provided in Section~\ref{sec:USC-MFGS in CL model} to such a larger class of model that we call the Generalized Caldeira-Leggett (GCL) model.
In order to proceed with the derivation of the USC-MFGS for such systems, we make two more physically motivated assumptions on the form of the potential $V_k(x,y)$ for our GCL model, i.e.,
\begin{align}
\lim_{x\to\pm\infty} V_k(\Tilde{x}_k(y) + x, y) &= \Omega(|x|) \quad \forall y \label{eqn:general CL model interaction hamiltonian assumption 1},\\
\frac{d}{dy} \Tilde{x}_k(y) &= \Omega(1) \label{eqn:requirement on Tilde{x}_k(x)}.
\end{align}
Here, we have used the big Omega notation \cite{knuth1976big} which, given two non-negative functions $f(x)$ and $g(x)$ such that
\begin{align}
    g(x) = \Omega(f(x)),
\end{align}
is defined to imply that there exists a positive number $C$ such that
\begin{align}
    g(x) \geq C f(x).
\end{align}
Also, here, $\Tilde{x}_k(y)$ is assumed to be continuous and differentiable everywhere. We remark that the continuity of the function $\Tilde{x}(y)$ is not affected by whether the system in question lives in a discrete dimensional Hilbert space or not because here we are treating $V_k(x,y)$ and $\Tilde{x}(y)$ as generic functions that can act on continuous variables as well as discrete operators.

Justification for Eq.~\ref{eqn:general CL model interaction hamiltonian assumption 1} is that keeping the system variable, $y$, fixed, if we move the environment variable $x$ far away from its global minima point $\Tilde{x}_k(y)$, then the value of the potential rises at least linearly with $|x-\Tilde{x}_k(y)|$. That is, we assume that the system and environment variables are attractively coupled to each other. 

Note that we can relax the assumption Eq.~\ref{eqn:requirement on Tilde{x}_k(x)} to $d \Tilde{x}_k(y)/dy  \geq \lambda$ for an arbitrarily small but finite $\lambda>0$, but we stick to the present form for simplicity.
We can motivate this assumption as following: if we instead assume that $\left|d\Tilde{x}_k(y) /dy \right| \approx 0$, then this would mean that for a large variation in $y$, $\Tilde{x}_k(y)$ would be almost constant (i.e., $\Tilde{x}_k(y)\approx x_0$) and for the corresponding value of the potential, we will have $V_k(x_0,y) \approx 0$ (Eq.~\ref{eqn:GCL minima point}). Hence, the assumption Eq.~\ref{eqn:requirement on Tilde{x}_k(x)} is applicable whenever the interaction potential is sensitive to a large change in the variable $y$ while $x$ is held constant.

The expression for the action, $S_k\left[q_k(t), A(t)\right]$ (Eq.~\ref{eqn:environment side path integral}) for the GCL model now becomes,
\begin{align}
S_k\left[q_k(t), A(t)\right] &= \int_0^\beta dt \bigg\{\frac{m_k}{2} \dot{q}_k(t)^2 + V_k(q_k(t), c_k A(t)) \bigg\}\label{eqn:GCL action}.
\end{align}
Eq.~\ref{eqn:general CL model interaction hamiltonian assumption 1} implies that, for a fixed value of $y$, there is a (perhaps) large but essentially finite basin with length scale $\Delta$ about the position $x = \Tilde{x}_k(y)$, such that all paths $q_k(t)$ that explore the region $|x - \Tilde{x}_k(y)| > \Delta$ will have a PE cost so high that the contribution of these paths to the path integral will be negligible. Hence, for all paths $q_k^*(t)$ that \textit{do} have a significant contribution to the path integral, we again have that
\begin{align}
| q_k^*(t) - \Tilde{x}_k(c_k A(t)) | < \Delta \label{eqn:valid bath paths for GCL model}.
\end{align}

Since $\Tilde{x}(c_k A(t))$ is ultimately a function of $t$, we can define $\braket{\Tilde{x}_A}$, $\braket{\Tilde{x}_A^2}$ and $\sigma_{\Tilde{x}_A}^2$ analogous to Eqs.~\ref{eqn:mean definition}-\ref{eqn: variance definition}.
Next, given Eq.~\ref{eqn:requirement on Tilde{x}_k(x)}, we prove in Appendix.~\ref{Appendix: GCL model proof} that
\begin{align}
\sigma_{\Tilde{x}_A}^2 \gtrsim c_k^2 \sigma_A^2 \label{eqn: sigma x is greater than c_k sigma a}.
\end{align}
Next, we note that if $\sigma_{\Tilde{x}_A} \approx \Delta$, then $\sigma_A = O(\Delta / c_k)$, which vanishes in the USC limit (Eq.~\ref{eqn:USC limit}). Alternatively, if we instead assume that
\begin{align}
\sigma_{\Tilde{x}_A}^2 \gg \Delta^2 , \label{eqn: GCL model condition KE PE proof}
\end{align}
then, using Eq.~\ref{eqn:valid bath paths for GCL model}, in Appendix.~\ref{Appendix: PE KE cost proof} we prove that
\begin{align}
\frac{1}{\beta}\int_0^\beta dt \dot{q}_k^*(t)^2 &\geq \mu \frac{\sigma_{\Tilde{x}_A}^2}{\beta^2}\\
&\gtrsim c_k^2 \mu \frac{\sigma_A^2}{\beta^2}, \quad \text{using Eq.~\ref{eqn: sigma x is greater than c_k sigma a}}\label{eqn: GCL variance conclusion}.
\end{align}
Here $\mu > 0$. We have already encountered an inequality similar to Eq.~\ref{eqn: GCL variance conclusion} in the context of the CL model (Eq.~\ref{eqn:kinetic energy cost}). Again, we are led to a similar conclusion that in the USC limit (Eq.~\ref{eqn:USC limit}), paths $q_k^*(t)$ (which have been defined to have convergent PE cost) will have divergent kinetic energy cost unless $\sigma_A$ vanishes. Therefore, again we have that $A(t) = A(0) = \Tilde{a}$, for some constant $\Tilde{a}$ (Eq.~\ref{eqn:frozen paths}). The expression for the influence functional (Eq.~\ref{eqn:environment side path integral}) hence simplifies as
\begin{align}
I_k[\Tilde{a}] &= \int_{-\infty}^{\infty} dx \int_{x}^{x} \mathcal{D}q_k(t) e^{-S_k\left[q_k(t), \Tilde{a}\right]}  \label{eqn:I_k path integral GCL}\\
\implies I[\Tilde{a}] &= \Tr \left[ e^{ -\beta \hat{H}_I(\Tilde{a})} \right].
\end{align}
Here, for a scalar $z$, we have defined 
\begin{align}
    \hat{H}_I(z) \equiv \sum_k\left[\hat{p}_k^2/(2 m_k) + V_k(\hat{q}_k, c_k z) \right].\label{eqn:Definition of HI(Ai)}
\end{align}
The expression for the full MFGS path integral (Eq.~\ref{eqn:discrete full path integral}) then becomes
\begin{align}
\hat{\rho}_{\text{USC}}(a, b) &= \Tilde{Z}^{-1} I[\Tilde{a}] \int_{a}^{b} \mathcal{D}\phi(t) \exp\left\{ -S_S[\Tilde{a}, \phi(t)] \right\}.
\end{align}
Here $a \text{ and }b$ are eigenvalues of the operator $\hat{\phi}$ and $\Tilde{a}$ is eigenvalue of $\hat{A}$. \cite{abandtilda} The USC-MFGS for GCL model is then given as
\begin{align}
\hat{\rho}_{\text{USC}} &= Z^{-1} \exp \left\{ -\beta \Tilde{H} \right\}\\
\Tilde{H} &\equiv \sum_i \hat{P}_i \left( \hat{H}_S + \hat{V}_0 \right) \hat{P}_i.
\end{align}
Here $Z$ is the normalization constant and $\hat{P}_i$ is the projection operator on the $i$th degenerate subspace of the operator $\hat{A}$ with eigenvalue $A_i$. Here, $\hat{V}_0$ is defined as
\begin{align}
\hat{V}_0 = \sum_i \log \left( \Tr\left( e^{-\beta \hat{H}_I(A_i)} \right) \right) \hat{P}_i, \label{eqn_definition_of_V0}
\end{align}
where $\hat{H}_I(A_i)$ is as defined in Eq.~\ref{eqn:Definition of HI(Ai)}.

The key takeaway here is that although the USC-MFGS for the GCL model is still diagonal in the basis of the system coupling-operator $\hat{A}$, the actual expression does deviate from the CL-USC result. Determining the actual expression for the GCL-USC result would involve calculating the Gibbs partition function corresponding to the free environment Hamiltonian, $\hat{H}_I(A_i)$, as a function of different values of $A_i$, and hence a closed form expression is not possible without further information on the form of $\hat{H}_I(A_i)$.

\subsection{Special Case} \label{sec:GCL2}
Since the CL-USC result gets modified for the GCL case, we will now identify a subclass within the GCL model, which we will call the GCL2 model, for which the USC-MFGS result remains the same as that for the CL case, and hence can also be determined in closed form.
To this end, let us assume a special form of the GCL model potential $V_k(x,y)$ (Eq.~\ref{eqn:more general interaction hamiltonian}) as
\begin{align}
V_k(x,y) &= U_k(x-y).
\end{align}
Generalization of the CL model of this form has been considered before in literature \cite{Wang_2019} and physically reflects a symmetry in the potential such that we have $V_k(x + w, y + w) = V_k(x,y)$. The function $\Tilde{x}_k(y)$, defined earlier, can now be expressed as
\begin{align}
\Tilde{x}_k(y) = y + x_k ,\label{eqn: x tilde for GCL2 model}
\end{align}
where $x_k$ is some constant. Note how $d\Tilde{x}_k(y)/dy  = 1$ and hence the corresponding GCL model assumption (Eq.~\ref{eqn:requirement on Tilde{x}_k(x)}) is automatically satisfied here. The requirement that $U_k(x)$ does not renormalize the free system Hamiltonian (Eq.~\ref{eqn:GCL minima point}), on the other hand, translates into
\begin{align}
U_k(x_k) &= 0 \label{eqn:x=y gives global minima}.
\end{align}
Similarly, the final GCL model assumption (Eq.~\ref{eqn:general CL model interaction hamiltonian assumption 1}) translates here as
\begin{align}
\lim_{x\to\pm\infty} U_k(x_k + x) &= \Omega(|x|) \label{eqn:diverging interaction potential for large distance}.
\end{align}

The remaining argument goes through just as before. The modified environment side action (Eq.~\ref{eqn:GCL action}),
\begin{align}
S_k[q_k(t), A(t)] \equiv \int_0^\beta dt \bigg\{ \frac{m_k}{2} \dot{q}_k(t)^2 \nonumber \\
\quad + U_k(q_k(t) - c_k A(t)) \bigg\},\label{eqn:GCL2 bath side action definition}
\end{align}
gives the following expression for the influence functional, analogous to Eq.~\ref{eqn:I_k path integral GCL},
\begin{align}
I_k[\Tilde{a}] &= \int_{-\infty}^{\infty} dx \int_{x}^{x} \mathcal{D}q_k(t) e^{-S_k[q_k(t), \Tilde{a}]}\\
&= \int_{-\infty}^{\infty} dx \int_{x-c_k\Tilde{a}}^{x-c_k\Tilde{a}} \mathcal{D}x_k(t) e^{-S_k[ x_k(t), 0]} \label{eqn:GCL2 case I_k calc step 2}\\
&= \int_{-\infty}^{\infty} dx \int_{x}^{x} \mathcal{D}x_k(t) e^{-S_k[x_k(t), 0]}\\
&\equiv I_k, \quad \text{which is independent of}\; \Tilde{a}.
\end{align}
Here, in Eq.~\ref{eqn:GCL2 case I_k calc step 2}, we defined $x_k(t) \equiv q_k(t) - c_k \Tilde{a}$ and we have used the relation $S_k\left[q_k(t), \Tilde{a}\right] = S_k\left[x_k(t), 0 \right]$ (Eq.~\ref{eqn:GCL2 bath side action definition}).

Hence, just like in the case of the CL model, we again have that $I[\Tilde{a}] \equiv I $ is independent of $\Tilde{a}$. This again leads to the familiar CL-USC result (Eq.~\ref{eqn:general USC state}). This means that CL-USC result is valid for a larger class of open system model, namely for GCL2 model, as defined here.

\section{USC-MFGS for Zwanzig model}\label{sec:beyond CL model}
We will now study the USC-MFGS for the so-called Zwanzig model \cite{zwanzig1973nonlinear, Banerjee_2002, Banerjee_2002_Kramers, Barik_2003}, which has been previously studied in the context of Brownian motion and barrier crossing problems. Let the full SE Hamiltonian for the Zwanzig model be given as $\hat{H}_{SE} = \hat{H}_S + \hat{H}_I$, where $\hat{H}_S$ is the free system Hamiltonian and $\hat{H}_I$ is given as
\begin{align}
\hat{H}_I &= \sum_k^N\left[ \frac{\hat{p}_k^2}{2 m_k} + U_k^{\text{free}}(\hat{q}_k) + \frac{c_k}{2} U_k(\hat{q}_k - \hat{A}) \right] \label{eqn:new spring model}.
\end{align}
Here $U_k^{\text{free}}(q_k)$ is an arbitrary free potential of the $k$th environment particle and $U_k(x)$ is as defined in Section~\ref{sec:GCL2}, but with just one extra condition that $U_k(x)$ now has a unique global minima point at $x=x_k$ and if there is any other local minima at a point $x=y_k$, then we have, for some non-infinitesimal positive number $\lambda$,
\begin{align}
U_k(y_k) - U_k(x_k) \geq \lambda > 0 \label{eqn:unique global minima}.
\end{align}
In other words, we now assume that not only does $U_k(x)$ have a unique global minimum, but also that if the potential has any other local minima, then there is a non-infinitesimal energy difference between the global minimum and the local minima. The significance of this assumption will become clear shortly.

Note that in Section~\ref{sec:GCL2}, the interaction potential was of the form $U_k(\hat{q}_k - c_k \hat{A})$ while here, it is of the form $c_k U_k(\hat{q}_k - \hat{A})$. Hence, the Zwanzig model is microscopically motivated to have a spring like interaction between the system and the environment particles. In fact, if $U_k(x)$ is a harmonic potential, then $c_k$ can be interpreted as the stiffness constant of the corresponding spring. Also note that here, we have deliberately not added a counter term in the interaction Hamiltonian because any renormalization of the system's free Hamiltonian will be assumed here to be of physical origin.

Now, the path integral expression for the MFGS will be given exactly as in CL model (Eq.~\ref{eqn:discrete full path integral} and Eq.~\ref{eqn:environment side path integral}), where the action associated with $I_k[A(t)]$ will now be given as 
\begin{align}
S_k[q_k(t), A(t)] \equiv \int_0^\beta dt \bigg\{ \frac{m_k}{2} \dot{q}_k(t)^2 + U_k^{\text{free}}(q_k(t))\nonumber \\
\quad + \frac{c_k}{2} U_k(q_k(t) - A(t))\bigg\}  \label{eqn:new model A(t)}.
\end{align}
In the USC limit (Eq.~\ref{eqn:USC limit}), for paths that contribute significantly to the path integral, the condition
\begin{align}
q_k(t) = A(t) + x_k \label{eqn:q_k = A new model}
\end{align}
will get strictly imposed here. This is because as discussed in the context of the GCL2 model, $U_k(x_k)=0$ is defined as the global minima of the potential $U_k(x)$ (Eq.~\ref{eqn:x=y gives global minima}) and we have further assumed that for Zwanzig model, this global minimum is unique and has a non-infinitesimal energy difference from any other local minima that might be there (Eq.~\ref{eqn:unique global minima}). Hence, in the USC limit, we will have
\begin{align}
\lim_{c_k \to \infty} \exp\left\{ - c_k U_k(x)\right\} = 
\begin{cases}
0 & \text{for } x \neq x_k \\
1 & \text{for } x = x_k
\end{cases}.
\end{align}
Replacing Eq.~\ref{eqn:q_k = A new model} in Eq.~\ref{eqn:new model A(t)}, the corresponding influence functional (Eq.~\ref{eqn:environment side path integral}) becomes
\begin{align}
I_k[A(t)] &= e^{-\int_0^\beta dt \left\{ \frac{m_k}{2} \dot{A}(t)^2 + U_k^{\text{free}}(A(t) + x_k) \right\}} \label{eqn:zwanzig influence functional}.
\end{align}
Since the influence functional $I_k[A(t)]$ (Eq.~\ref{eqn:environment side path integral}) arises when we take partial trace over the environment degrees of freedom, the environment side paths, $q_k(t)$, come with a constraint that $q_k(0) = q_k(\beta)$. In the USC limit, since $q_k(t)$ gets further constrained as $q_k(t) = A(t) + x_k$ (Eq.~\ref{eqn:q_k = A new model}), this in turn puts a constraint that for paths that contribute to the path integral, $A(0) = A(\beta)$. This effectively diagonalizes the final USC-MFGS in the basis of system coupling operator $\hat{A}$, but through a mathematical mechanism that is different from the one encountered in the GCL model.

\subsection{Special Case 1: Discrete system}\label{sec:zwanzig discrete case}
If the system lives in a discrete Hilbert space, $A(t)$ generally has instantaneous jumps, in which case the kinetic term in the action in $I_k[A(t)]$ will blow up (Eq.~\ref{eqn:zwanzig influence functional}), preventing the jump. Hence, for a discrete system, paths with an instantaneous jump will be suppressed and we will have $\dot{A}(t) = 0$.

Next, for $A(t) = \Tilde{a}$, the influence functional (Eq.~\ref{eqn:zwanzig influence functional}) simplifies as
\begin{align}
\implies  I_k[\Tilde{a}] &= \exp \left\{-\int_0^\beta dt U_k^{\text{free}}(\Tilde{a} + x_k)  \right\}\\
&= \exp \left\{-\beta U_k^{\text{free}}(\Tilde{a} + x_k)  \right\}.
\end{align}
Replacing this in the full MFGS path integral expression (Eq.~\ref{eqn:discrete full path integral}) gives
\begin{align}
\hat{\rho}_{\text{USC}}(a,b) &= \Tilde{Z}^{-1} e^{-\beta U_{\text{eff}}(\Tilde{a})   }\int_{a}^{b} \mathcal{D}\phi(t) \exp \left\{-S_S[\Tilde{a}, \phi(t)] \right\}.
\end{align}
Here $a \text{ and }b$ are the eigenvalues of the operator $\hat{\phi}$ and $\Tilde{a}$ is the eigenvalue \cite{abandtilda} of $\hat{A}$ and $U_{\text{eff}}(x)$ is defined as
\begin{align}
U_{\text{eff}}(x) \equiv \sum_{k=1}^N U_k^{\text{free}}(x + x_k).
\end{align}
Finally, the USC-MFGS is given as,
\begin{align}
\implies \hat{\rho}_{\text{USC}} &= Z^{-1} \exp \left\{ -\beta \Tilde{H} \right\} \label{eqn:Zwanzig model discrete case USC-MFGS}\\
\Tilde{H} &\equiv \sum_i \hat{P}_i (\hat{H}_S + \hat{V}_0) \hat{P}_i \label{eqn:beyond cl hmf}.
\end{align}
Here $\hat{P}_i$ is the projection operator on the $i$th degenerate subspaces of the operator $\hat{A}$ with eigenvalue $A_i$. Here $\hat{V}_0$ is defined as
\begin{align}
\hat{V}_0 = \sum_i U_{\text{eff}}(A_i) \hat{P}_i.
\end{align}
Hence, effectively, this is just the USC-MFGS result for GCL model, but with a renormalized free system Hamiltonian (Eq.~\ref{eqn:beyond cl hmf}).
\begin{figure}[htbp]
\includegraphics[width=0.48\textwidth]{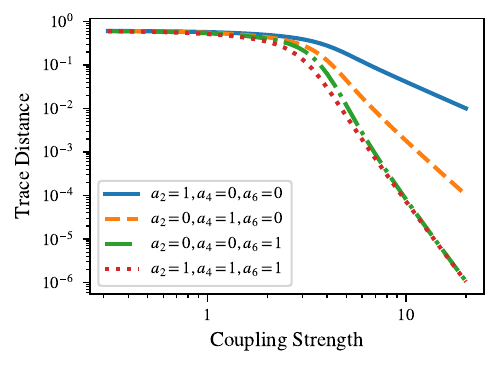}
\caption{ The trace distance between the USC-MFGS for the GCL2 model and numerically evaluated MFGS as a function of the coupling strength $c$, for different choices of the interaction potential (Eq.~\ref{eqn:GCL2 interaction potential}).}
\label{fig:qutrit_ho_usc}
\end{figure}

\subsection{Special Case 2: Continuous Variable system}
Now, assume that we are dealing with a CV system, where we have a particle with mass $m$ in a potential $V(q)$. Then, the free Hamiltonian $\hat{H}_S$ is given as
\begin{align}
\hat{H}_S &= \frac{\hat{p}^2}{2m} + V(\hat{q}).
\end{align}
Also, for simplicity, assume that the system coupling-operator is given as $\hat{A} = \hat{q}$. Then, using the simplified expression for the influence functional for Zwanzig model (Eq.~\ref{eqn:zwanzig influence functional}) in the full MFGS path integral (Eq.~\ref{eqn:discrete full path integral}) gives the USC-MFGS as
\begin{align}
\hat{\rho}_{\text{USC}}(x,y) &= \Tilde{Z}^{-1} \delta(x,y) \int_{x}^{x} \mathcal{D}q(t)\nonumber \\
& \quad \exp\left\{-\int_0^\beta dt  \left( \frac{1}{2}M_{\text{eff}} \dot{q}(t)^2  + V_{\text{eff}}(q(t)) \right) \right\} 
\end{align}
where,
\begin{align}
M_{\text{eff}} &= m + \sum_k m_k\\
V_{\text{eff}}(x) &= V(x) + U_{\text{eff}}(x)
\end{align}
The USC-MFGS can now be determined as
\begin{align}
\braket{ q | \rho_{\text{USC}} | q'} &= Z^{-1} \delta(q,q') \braket{ q |e^{-\beta \hat{H}_{\text{eff}}} | q'} \label{eqn:zwanzig USC result},\\
\hat{H}_{\text{eff}} &= \frac{\hat{p}^2}{2 M_{\text{eff}}} + V_{\text{eff}}(\hat{q}).
\end{align}

Note one key distinction here from the GCL2-USC result, in that although both the states are diagonal in the basis set by the system coupling-operator, the corresponding GCL2-USC result for CV system would be of the form, $\rho_{\text{USC-GCL2}} = Z^{-1} e^{-\beta V_{\text{eff}}(q)}$, as derived in Appendix.~\ref{appendix_CL_USC_result_for_CV_system}. \cite{ankerhold2003phase, cresser2021weak, kumar2024local} The Zwanzig-USC result (Eq.~\ref{eqn:zwanzig USC result}) will converge to the GCL2 result only when $M_{\text{eff}} \to \infty$, which will suppress paths of the type $\dot{q}(t) \neq 0$ in the path integral. This will typically not be the case for a system interacting with a few environment particles or particles that are not very massive, for example.

We note that there has been a disagreement in the literature regarding the form of the USC equilibrium state under the CL model.\cite{cresser2021weak} For the CL model, Kawai et al. conjectured \cite{goyal2019steady, orman2020qubit} the system's dynamical state at long times to be $\rho^{\text{conj}} = \sum_i \hat{P}_i e^{-\beta \hat{H}_0} \hat{P}_i$ and we note that, coincidentally, the Zwanzig-USC result for CV systems (Eq.~\ref{eqn:zwanzig USC result}) is similar to this conjectured form.

\section{Numerical verification} \label{sec:Numerical verification}
\begin{figure}[htbp]
\includegraphics[width=0.48\textwidth]{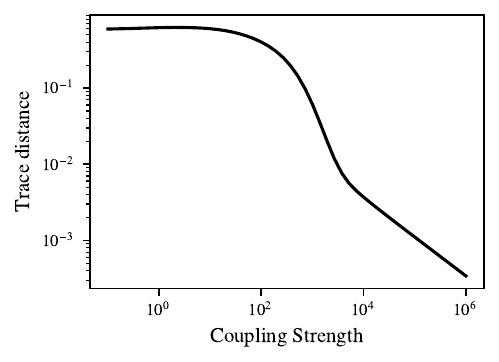}
\caption{ The trace distance between the USC-MFGS for the Zwanzig model and the numerically calculated MFGS as a function of the coupling strength $c$.}
\label{fig:qutrit_deviation_from_cresser_anders}
\end{figure}
In this section, we conduct numerical tests to validate the analytical results presented in this paper. Specifically, we verify the USC-MFGS results for both the GCL2 (Eq.~\ref{eqn:general USC state}) and the Zwanzig model (Eq.~\ref{eqn:Zwanzig model discrete case USC-MFGS}). Additionally, we examine how the convergence rate of the MFGS towards the USC results varies with an increase in the SE coupling, across different model parameters.

For both the GCL2 and Zwanzig model, we consider a simplified scenario involving a qutrit system interacting with a single environmental particle. This allows for the convenient numerical evaluation of the exact MFGS, denoted by $\hat{\rho}_{\text{num}}$ (see Appendix.~\ref{appendix_numerical_details} for numerical details). 
By increasing the coupling strength, we can then compare the approach of $\hat{\rho}_{\text{num}}$ to the corresponding $\hat{\rho}_{\text{USC}}$ derived in this paper. 
We note that the system MFGS will get closer to the USC-result if multiple environment particles are simultaneously made to interact it.

The system free Hamiltonian, $\hat{H}_S$, and the system coupling-operator, $\hat{A}$, are fixed throughout to be
\begin{align} \label{eqn_system_free_hamiltonian_num}
\hat{H}_S &= \begin{pmatrix}
1 & 1 & 0 \\
1 & 0 & 1 \\
0 & 1 & -1
\end{pmatrix}, \quad
\hat{A} = \begin{pmatrix}
1 & 0 & 0 \\
0 & 0 & 0 \\
0 & 0 & -0.5
\end{pmatrix}.
\end{align}
We set $\beta = 5 E_h$ throughout, where we are working in the Hartree atomic units.
For the GCL model, the interaction Hamiltonian, $\hat{H}_I$, (Eq.~\ref{eqn:more general interaction hamiltonian}) simplifies as
\begin{align}
\hat{H}_I &= \frac{\hat{P}^2}{2 M} + V(\hat{Q} - c \hat{A}). \label{eqn_GCL_num_H}
\end{align}
Here $M=1$ and $V(x)$ is assumed to be of the form
\begin{align}
V(x) &= \sum_{n=1}^3 a_{2n} x^{2n} \label{eqn:GCL2 interaction potential}.
\end{align}

Fig.~\ref{fig:qutrit_ho_usc} plots the trace distance between $\hat{\rho}_{\text{num}}$ and $\hat{\rho}_{\text{USC}}$ (Eq.~\ref{eqn:general USC state}) for this model as we increase the coupling parameter, $c$, for different values of the positive semi-definite coefficients $a_{2n}$. We generically find that, as expected, the trace distance drops quickly as the value of $c$ is increased. We also find that higher order SE coupling causes the MFGS to approach the USC state faster as a function of the coupling parameter $c$.

This behavior can be explained for a generic GCL model if we note that a path $A(t)$ will contribute significantly to the path integral only if we have $\sigma_A = O(\Delta/c_k)$. For USC limit, we therefore require that $\Delta/c_k \ll 1$. Hence, the USC limit is approached based on the interplay between the coupling parameter $c_k$ and $\Delta$. \cite{deltaexplaination} So, if $\Delta$ is small for some coupling potential $V_k(x,y)$, like in the case of the potentials with $a_4 = 1$ or $a_6 = 1$ in Fig.~\ref{fig:qutrit_ho_usc}, then it will facilitate the approach to the USC limit.

Next, for Zwanzig model, the interaction Hamiltonian $\hat{H}_I$ (Eq.~\ref{eqn:new spring model}) is assumed to be
\begin{align}
\hat{H}_I &=  \frac{\hat{P}^2}{2 M} +  U(\hat{Q}) + \frac{c}{2} (\hat{Q}-\highlighttext{\hat{A}})^2. \label{eqn_zwanzig_num_H}
\end{align}
Here, again, $M=1$ and $U(x)$ is a Morse potential
\begin{align}
U(x) \equiv (1 - e^{-x})^2.
\end{align}

Fig.~\ref{fig:qutrit_deviation_from_cresser_anders} plots the trace distance between $\hat{\rho}_{\text{num}}$ and $\hat{\rho}_{\text{USC}}$ (Eq.~\ref{eqn:Zwanzig model discrete case USC-MFGS}) for this model as we increase the value of the coupling parameter, $c$. We again observe that, as expected, the trace distance drops quickly as the value of $c$ is increased, though the rate of approach to the USC result is much slower than the GCL case.

\section{Conclusion} \label{sec:Conclusion}
The MFGS result, defined in Eq.~\ref{eqn:definition of MFGS}, has been studied in USC limit for a system coupled to an anharmonic environment, hence going beyond the known USC result valid for the CL model. For all the generalized SE models studied, the USC-MFGS has been found to always be diagonal in the basis set by the SE interaction, although the exact form of the state differs from the CL-USC result, like in the case of a quite general class of models that we call the GCL model (Section~\ref{sec:generalization 1}). We also identify a subclass of the GCL model, which we call the GCL2 model (Section~\ref{sec:GCL2}), for which the USC-MFGS remains equivalent to the CL-USC result. Furthermore, for a so-called Zwanzig model (distinct from the GCL model) (Section~\ref{sec:beyond CL model}), analytical expression for the USC state has been derived and is found to have qualitative differences from the GCL-USC result (Eq.~\ref{eqn:zwanzig USC result}).

These results demonstrate how the CL-USC result is influenced when the different assumptions of the model for which it was derived are relaxed one by one.
In particular, these results directly extend the applicability of the CL-USC result for several strong coupling quantum thermodynamics, chemistry and biological problems where the anharmonicity of the environment becomes relevant. Also, our results indicate that anharmonicity of the environment has important implications on the rate of approach of the MFGS to the USC limit as a function of the coupling strength (Fig.~\ref{fig:qutrit_ho_usc}). As a future direction, it would be interesting to apply our  results to study some systems of physical interest (see for example \cite{wang2007quantum, wang2004semiclassical}).

The CL model has served as an important tool to model a variety of physical processes \cite{feynman2000theory, Leggett1987dynamics, gilmore2006criteria, nazir2009correlation, Xu_1994, Ishizaki_2009, Rebentrost_2009, de_Vega_2011, Mohseni_2013, Tang_1993, Barbara_1996, Mavros_2016, Zhao_2012, Tanimura_1993, weiss2012quantum, BRE02, gardiner2004quantum}, and its strengths and limitations in capturing the properties of these systems of interest have also been investigated. \cite{bramberger2020dephasing, rognoni2021caldeira, gottwald2015applicability} Future research into extending such studies beyond the CL model (for example, to GCL2 and the Zwanzig model) will be of high interest. Moreover, the lines of reasoning used in this paper to obtain the USC-MFGS result can be applied to study USC-MFGS for other physically motivated generalized SE models.

Our results across a broad range of models indicate that the USC-MFGS invariably becomes diagonal in the basis of the system coupling-operator. Looking forward, it would be interesting to investigate whether this diagonalization is a universal feature of the USC-limit, independent of the further details of the SE model.
Additionally, the CL-USC result has previously been expanded to first order in inverse of the SE coupling \cite{Quanta167} and has also been approximated for intermediate and weak coupling regime. \cite{anto2023effective, anto2023effective2} Extending these results beyond the CL model would be a valuable direction for future research.

\section*{AUTHOR DECLARATIONS}

\subsection*{Conflict of Interest}

The authors have no conflicts to disclose.

\subsection*{Author Contributions}

\textbf{Prem Kumar}: Methodology (equal); Validation (equal); Formal analysis (equal); Investigation (equal); Conceptualization (equal); Writing – original draft (equal); Writing – review \& editing (equal). \textbf{Sibasish Ghosh}: Methodology (equal); Validation (equal); Formal analysis (equal); Investigation (equal); Conceptualization (equal); Writing – original draft (equal); Writing – review \& editing (equal). 

\section*{DATA AVAILABILITY}

The data that support the findings of this study are available within the article.

\begin{appendix}

\section{Proposition 1}\label{Appendix: PE KE cost proof}
\begin{mybox}[]
Let $f(t)$ be a continuous, differentiable and square integrable function such that its derivative, $\dot{f}(t)$, is also square integrable, such that
\begin{align}
f(t) \equiv g(t) + \epsilon(t) \label{eqn:defining eqn}.
\end{align}
Here $g(t)$ is also square integrable. Next, for some $\Delta > 0$, we have the constraint that
\begin{align}
| \epsilon(t) | < \Delta \label{eqn: limit on epsilon}.
\end{align}
Next, let us define
\begin{align}
\braket{g} &\equiv \frac{ 1 }{\beta}\int_0^\beta dt g(t),\\
\braket{g^2} &\equiv \frac{ 1 }{\beta} \int_0^\beta dt g(t)^2,\\
\sigma_g^2 &\equiv \braket{g^2} - \braket{g}^2 \label{eqn: definition of sigma_g}.
\end{align}
Then, in the limit $\sigma_g^2 \gg \Delta^2$, we will prove that
\begin{align}
\frac{1}{\beta}\int_0^\beta dt \dot{f}(t)^2 &\geq \mu \frac{\sigma_g^2}{\beta^2} \label{eqn: to prove}, %my guess about one less factor of \beta
\end{align}
where $\mu > 0$ is a constant.
\end{mybox}

The intuitive idea behind the proof is as following. Part 1 of the proof consists of proving that
\begin{align}
\lim_{\sigma_g^2 \gg \Delta^2}\sigma_f^2 \approx \sigma_g^2 \label{eqn:proof1 part 1}.
\end{align}
Part 2 then consists of proving that given $\sigma_f$, 
\begin{align}
\frac{1}{\beta}\int_0^\beta dt \dot{f}(t)^2 \geq \mu \sigma_f^2/\beta^2, \quad \text{where }\mu >0\label{eqn:proof1 part 2}.
\end{align}
Replacing Eq.~\ref{eqn:proof1 part 1} into Eq.~\ref{eqn:proof1 part 2} then gives the final result (Eq.~\ref{eqn: to prove}).

Both these claims are intuitive, as the first claim states that if the absolute difference between two functions is much smaller than the standard deviation of one of the functions, then their standard deviations are also approximately equal. The second claim states that, for a function $f(t)$, if $\sigma_f^2>0$, then $\braket{\dot{f}^2} >0$ and also that $\braket{\dot{f}^2} \propto \sigma_f^2$, which makes sense because if you scale a function as $f(t) \to 2 f(t)$, then we have $\sigma_f^2 \to 4 \sigma_f^2$ and $\braket{\dot{f}^2} \to 4 \braket{\dot{f}^2}$.

\textbf{Part 1 of the proof:} Let us calculate $\sigma_f^2$ (using Eq.~\ref{eqn:defining eqn}) as
\begin{align}
\sigma_f^2 &= \braket{(g + \epsilon)^2} - \braket{g + \epsilon}^2\\
&= \braket{g^2} + 2 \braket{ g \epsilon} + \braket{\epsilon^2} - \left( \braket{g}^2 + 2 \braket{g} \braket{\epsilon} + \braket{\epsilon}^2 \right)\\
&= \sigma_g^2 + \sigma_\epsilon^2 + 2 \left( \braket{g \epsilon} - \braket{g} \braket{\epsilon} \right)\\
&\leq \sigma_g^2 + O(\Delta^2) + 2 \sigma_g O(\Delta).
\end{align}
Here, we used $\braket{g \epsilon} - \braket{g} \braket{\epsilon} \leq \sigma_g \sigma_\epsilon$ that can be derived using the Cauchy-Schwarz inequality. Finally, in the limit $\sigma_g^2 \gg \Delta^2$, we get the required result (Eq.~\ref{eqn:proof1 part 1}).

\textbf{Part 2 of the proof:} Proof of Eq.~\ref{eqn:proof1 part 2} follows directly from Poincaré inequality \cite{leoni2024first, evans2022partial} that can be formally stated as following.\cite{ stover_poincare}
\begin{mybox}[Poincaré inequality]
Let $\Omega$ be an open, bounded, and connected subset of $\mathbb{R}^d$ for some $d$ and let $d \mathbf{x}$ denote $d$-dimensional Lebesgue measure on $\mathbb{R}^d$. Then the Poincaré inequality says that there exist constants $C_1$ and $C_2$ such that
\begin{align}
\int_\Omega h^2(\mathbf{x}) d \mathbf{x} &\leq C_1 \int_\Omega | \nabla h(\mathbf{x})|^2 d \mathbf{x} \nonumber\\
& \quad + C_2 \left[ \int_\Omega h(\mathbf{x}) d \mathbf{x} \right]^2,
\end{align}
for all functions $h$ in the Sobolev Space $H^1(\Omega)$ consisting of all functions in $L^2(\Omega)$ whose generalized derivatives are all also square integrable.
\end{mybox}

For 1-dimensional Euclidean space, for simplicity and without loss of any generality, let us redefine $f(t)$ such that $\braket{f} = 0$. In other words, we shift $f(t) \to f(t) - \braket{f}$. Then, the Poincaré inequality implies that
\begin{align}
\braket{f^2} &\leq C_1 \braket{\dot{f}^2} + \beta C_2 \braket{f}^2\\
\implies \sigma_f^2 &\leq C_1 \braket{\dot{f}^2}.
\end{align}
Since $\sigma_f^2 \geq 0$ and $ \braket{\dot{f}^2} \geq 0$, we have that $C_1$ is positive. We therefore have
\begin{align}
\braket{\dot{f}^2} \geq \frac{\beta^2}{C_1} \frac{\sigma_g^2}{\beta^2}.
\end{align}
This proves Eq.~\ref{eqn:proof1 part 2} for $\mu = \beta^2/C_1$ and concludes the proof of the main result (Eq.~\ref{eqn: to prove}) as well.

\subsection{Proof of special case of Poincaré Inequality}

Here, for completeness, we provide a proof for the special case of Poincaré inequality (Eq.~\ref{eqn:proof1 part 2}), as per the present requirement. To simplify the proof, other than assuming that $\braket{f}=0$, we also shift the range of the imaginary time from the conventional $0 \leq t \leq \beta$ to $-\beta/2 \leq t \leq \beta/2$. Now, given a constraint
\begin{align}
\int_{-\frac{\beta}{2}}^{\frac{\beta}{2}} f(t)^2 dt = \beta \sigma_f^2 \label{eqn:constraint equation},
\end{align}
we need to minimize over the action
\begin{align}
S = \int_{-\frac{\beta}{2}}^{\frac{\beta}{2}} dt \dot{f}(t)^2 \label{eqn: kinetic action}.
\end{align}
We do this by perturbing the optimum path by a small amount, $f(t) \to f(t) + \delta(t)$, with $\delta(-\beta/2) = \delta(\beta/2) = 0$. Then Eq.~\ref{eqn:constraint equation} demands that to the first order in $\delta(t)$, we have
\begin{align}
\int_{-\frac{\beta}{2}}^{\frac{\beta}{2}} f(t) \delta(t) dt = 0 \label{eqn:contrained perturbation}.
\end{align}
Now, the condition for the minimum action is $\delta S = 0$, where
\begin{align}
\frac{\delta S}{2} &= \int_{-\frac{\beta}{2}}^{\frac{\beta}{2}} dt \dot{f}(t)\dot{\delta}(t)\\
&= \dot{f}(\beta/2) \delta(\beta/2) - \dot{f}(-\beta/2) \delta(-\beta/2)\nonumber\\
& \quad \quad - \int_{-\frac{\beta}{2}}^{\frac{\beta}{2}} dt \ddot{f}(t)\delta(t)\\
&= -\int_{-\frac{\beta}{2}}^{\frac{\beta}{2}} dt \ddot{f}(t)\delta(t) = 0\label{eqn:equation for minimum action}.
\end{align}
Comparing Eq.~\ref{eqn:equation for minimum action} with Eq.~\ref{eqn:contrained perturbation}, for some constant $\lambda$, we get
\begin{align}
\ddot{f}(t) &= \pm \lambda^2 f(t).
\end{align}
The solution of $f(t)$ can be of either sinusoidal or hyperbolic type. For the sinusoidal solution, we have
\begin{align}
f(t) = A \sin(\lambda t + \phi).
\end{align}
Then the condition, $\braket{f} = 0$, demands that
\begin{align}
\int_{-\frac{\beta}{2}}^{\frac{\beta}{2}} A \sin(\lambda t + \phi) &= 0\\
\implies -\frac{A}{\lambda} \left(\cos\left(\frac{\lambda \beta}{2} + \phi\right) - \cos\left(-\frac{\lambda \beta}{2} + \phi\right)\right) &= 0\\
\implies \cos\left(\frac{\lambda \beta}{2} + \phi\right) = \cos\left(-\frac{\lambda \beta}{2} + \phi\right)&.
\end{align}
The solution is either, for a generic $\lambda$, $\phi = 0$ or for a generic $\phi$, $\lambda = 2 \pi n/\beta$ for $n\in \mathbb{Z}$. Hence, we have
\begin{align}
f(t) = 
\begin{cases}
A \sin(\lambda t) \quad \text{ for generic } \lambda\\
A \sin(\lambda t + \phi) \quad \text{ for }\lambda = 2 \pi n/\beta; \quad n \in \mathbb{Z}\\
\end{cases}.
\end{align}

For hyperbolic solution, $f(t)$ is given as
\begin{align}
f(t) = A \sinh(\lambda t).
\end{align}
Here, we do not have a $\cosh$ kind of term because of the requirement that $\braket{f}= \int_{-\beta/2}^{\beta/2} dt f(t) = 0$.

\subsubsection{Sinusoidal solution}
Let us first consider the solution for $\phi = 0$ and a generic $\lambda$
\begin{align}
f(t) &= A \sin(\lambda t)\label{eqn:sinusoidal phi=0 case}\\
\implies \beta \braket{f^2} &= A^2 \int_{-\frac{\beta}{2}}^{\frac{\beta}{2}} \sin^2(\lambda t) dt\\
&= \frac{A^2}{2} \int_{-\frac{\beta}{2}}^{\frac{\beta}{2}} 1 - \cos(2\lambda t) dt\label{eqn: sinusoidal 3rd line}\\
&= \frac{A^2}{2} \left( \beta - \frac{2 \sin(\lambda \beta)}{2 \lambda} \right) \\
&= \frac{A^2 \beta}{2} \left( 1 - \frac{\sin(x)}{x} \right) \quad ;x \equiv \lambda \beta\\
\implies A^2 &= \braket{f^2}  \frac{ 2x}{x- \sin(x)}\label{eqn:sinusoidal A^2}.
\end{align}

\begin{figure}[htbp]
\includegraphics[width=0.48\textwidth]{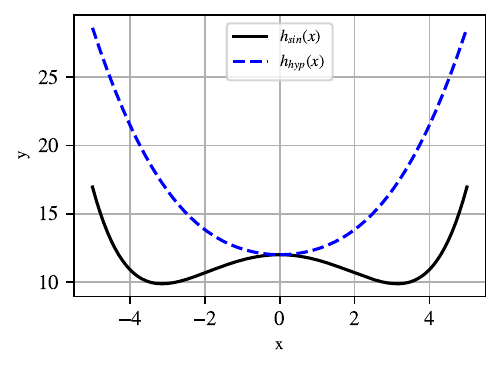}
\caption{ Plot of $h_{\text{sin}}(x)$ (Eq.~\ref{eqn: h sin x}) and $h_{\text{hyp}}(x)$ (Eq.~\ref{eqn: h hyp x}), which demonstrates that the minima of both these functions are larger than zero.}
\label{fig:sine_hyperbolic}
\end{figure}

This expression for $A^2$ is crucial, so let us also evaluate it for the case when, in Eq.~\ref{eqn:sinusoidal phi=0 case}, $\phi$ is present and $\lambda = 2 \pi n/\beta$ for $n\in \mathbb{Z}$.
\begin{align}
f(t) &= A \sin\left(\frac{2 \pi n t}{\beta} + \phi\right)\\
\implies \beta \braket{f^2} &= A^2 \int_{-\frac{\beta}{2}}^{\frac{\beta}{2}} \sin^2\left(\frac{2 \pi n t}{\beta} + \phi\right) dt\\
&= \frac{A^2 \beta}{2}\\
\implies A^2 &= 2 \braket{f^2}\\
&= \braket{f^2} \frac{ 2x}{x- \sin(x)} \bigg{|}_{x = 2 \pi n}
\end{align}
This is consistent with Eq.~\ref{eqn:sinusoidal A^2}. Hence, Eq.~\ref{eqn:sinusoidal A^2} holds generically for the full sinusoidal case. Next, we have that
\begin{align}
f(t)^2 + \frac{\dot{f}(t)^2}{\lambda^2} &= A^2\\
\implies \braket{f^2} + \frac{\braket{\dot{f}^2}}{\lambda^2} &= A^2
\end{align}
\begin{align}
\implies \braket{\dot{f}^2} &= \lambda^2 \left( A^2 - \braket{f^2} \right) \label{eqn:sinusoidal step 3}\\
&= \frac{x^2}{\beta^2} \left( \braket{f^2} \frac{2 x}{x- \sin(x)}- \braket{f^2} \right) \label{eqn:sinusoidal step 4}\\
&= \frac{\braket{f^2} }{\beta^2} x^2 \left( \frac{x + \sin(x)}{x - \sin(x)}\right)\\
&\geq \left(\min_x h_{\text{sin}}(x) \right) \frac{\braket{f^2} }{\beta^2} \label{eqn:sinusoidal solution}.
\end{align}
Here
\begin{align}
h_{\text{sin}}(x) &\equiv x^2 \left( \frac{x + \sin(x)}{x - \sin(x)}\right) \label{eqn: h sin x}.
\end{align}
Note that $\min_{x} h_{\text{sin}}(x) > 0$, which can also be seen from Fig.~\ref{fig:sine_hyperbolic}. Eq.~\ref{eqn:sinusoidal solution} is the desired result for the sinusoidal case.

\subsubsection{Hyperbolic solution}
Let
\begin{align}
f(t) &\equiv A \sinh(\lambda t)\\
\beta \braket{f^2} &\equiv A^2 \int_{-\frac{\beta}{2}}^{\frac{\beta}{2}} \sinh^2(\lambda t) dt\\
&= - \frac{A^2}{2} \int_{-\frac{\beta}{2}}^{\frac{\beta}{2}} 1 - \cosh(2\lambda t) dt\\
\implies A^2 &= - \braket{f^2} \frac{2x}{x - \sinh(x)}.
\end{align}
Here, in the last step, we used the analogous derivation from Eq.~\ref{eqn: sinusoidal 3rd line} to Eq.~\ref{eqn:sinusoidal A^2}, but with an extra minus sign. Now, note that we have
\begin{align}
\frac{\dot{f}(t)^2}{\lambda^2} - f(t)^2 &= A^2\\
\implies \frac{\braket{\dot{f}^2}}{\lambda^2} - \braket{f^2} &= A^2
\end{align}
\begin{align}
\implies \braket{\dot{f}^2} &= -\lambda^2 \left( -A^2 - \braket{f^2} \right) \label{eqn:hyperbolic step 3}\\
&= -\frac{x^2}{\beta^2} \left( \braket{f^2} \frac{2x}{x - \sinh(x)} - \braket{f^2} \right)\label{eqn:hyperbolic step 4}\\
&\geq \left(\min_x h_{\text{hyp}}(x) \right) \frac{\braket{f^2} }{\beta^2} \label{eqn:hyperbolic solution}.
\end{align}
In the last step, we used the analogous calculation from Eq.~\ref{eqn:sinusoidal step 4} to Eq.~\ref{eqn:sinusoidal solution}, with just an extra minus sign. Here, $h_{\text{hyp}}(x)$ is defined as
\begin{align}
h_{\text{hyp}}(x) &\equiv x^2 \left( \frac{\sinh(x) + x}{\sinh(x) - x}\right)\label{eqn: h hyp x}.
\end{align}

Note that $\min_{x} h_{\text{hyp}}(x) > 0$, which can also be seen from Fig.~\ref{fig:sine_hyperbolic}. Eq.~\ref{eqn:hyperbolic solution} is the desired result for the hyperbolic case.

\section{Proposition 2} \label{Appendix: GCL model proof}
\begin{mybox}[]
Let $f(x)$ be a continuous, differentiable and square integrable function and let us define $h(x) \equiv \frac{d}{dx} f(x)$ such that
\begin{align}
h(x)  = \Omega(1) \label{eqn: constraint on the derivation of h(x)}.
\end{align}
Now, let $g(t)$ be another square integrable function and
\begin{align}
\braket{f \circ g} &\equiv \frac{ 1 }{\beta} \int_0^\beta dt f(g(t)) \label{eqn_fog_avg_defn}\\
\braket{(f \circ g)^2} &\equiv \frac{ 1 }{\beta} \int_0^\beta dt f(g(t))^2\\
\sigma_{f \circ g}^2 &\equiv \braket{(f \circ g)^2} - \braket{f \circ g}^2.
\end{align}
Then, we will prove that
\begin{align}
\sigma_{f \circ g}^2 \gtrsim \sigma_g^2,
\end{align}
where $\sigma_g^2$ is defined analogously (Eq.~\ref{eqn: definition of sigma_g}).
\end{mybox}

\textbf{Proof:} 
We first write
\begin{align}
f(x_1) &= f(x_0) + \int_{x_0}^{x_1} h(x') dx'.
\end{align}
Now, from Eq.~\ref{eqn: constraint on the derivation of h(x)}, if $x_1 > x_0$, then
\begin{align}
\int_{x_0}^{x_1} h(x') dx' &\gtrsim x_1 - x_0 > 0\\
\implies \int_{x_1}^{x_0} h(x') dx' &\lesssim x_0 - x_1 < 0\\
\implies \left| \int_{a}^{b} h(x') dx' \right| &\gtrsim |b - a| \quad \forall a,b\\
\implies |f(b) - f(a)| &\gtrsim |b - a|.
\end{align}
Then we have
\begin{align}
|(f \circ g)(t_1) - (f \circ g)(t_2)| &\gtrsim |g(t_1) - g(t_2)| \quad \forall t_1, t_2.
\end{align}

Now, let $t'$ be the time at which we have $(f \circ g)(t') = \braket{f \circ g}$. Then we have
\begin{align}
\sigma_{f \circ g}^2 &= \frac{ 1 }{\beta} \int_0^\beta dt ((f \circ g)(t) - (f \circ g)(t'))^2\\
&\gtrsim \frac{ 1 }{\beta} \int_0^\beta dt (g(t) - g(t'))^2\\
&= \frac{1}{\beta}  \int_0^\beta dt (g(t) - (\braket{g} + c))^2 \text{ where } c \equiv g(t') - \braket{g} \label{eqn_appendix_b_c_definition}\\
&= \frac{1}{\beta} \int_0^\beta dt \left[ (g(t) - \braket{g})^2 + c^2 - 2 c (g(t) - \braket{g})\right]\\
&= \sigma_g^2 + c^2 -2c (\braket{g} - \braket{g})\\
&\geq \sigma_g^2\\
&\implies \sigma_{f \circ g}^2 \gtrsim \sigma_g^2.
\end{align}

\section{Numerical evaluation of the MFGS for qutrit system coupled to single anharmonic oscillator at finite SE coupling} \label{appendix_numerical_details}

The Hamiltonian of a qutrit system coupled to a single anharmonic oscillator (Eq.~\ref{eqn_system_free_hamiltonian_num} together with Eq.~\ref{eqn_GCL_num_H} (for GCL2 model) and Eq.~\ref{eqn_zwanzig_num_H} (for Zwanzig-model)) is essentially an infinite-dimensional operator. In order to numerically evaluate the MFGS,
\begin{align}
\hat{\rho} = Z^{-1}\Tr_{E}\left[ e^{-\beta \hat{H}_{\text{SE}}}\right], \label{eqn_appendix_mfgs_num}
\end{align}
we truncate the position operator $\hat{Q}$ of environment oscillator so that its eigenvalues $Q$ is restricted to $Q_{min} \leq Q \leq Q_{max}$. We also discretize $\hat{Q}$ into $N$ discrete points. This effectively approximates the original infinite-dimensional Hamiltonian $\hat{H}_{\text{SE}}$ to one with $3N$ dimensions.

It is now possible to numerically evaluate the MFGS (Eq.~\ref{eqn_appendix_mfgs_num}). We vary the values of $Q_{min}, Q_{max}$ and $N$ until the final result of interest converges (in this case, the trace distance between $\hat{\rho}_{\text{num}}$ and $\hat{\rho}_{USC}$).

\section{USC-MFGS for CV system for CL model}\label{appendix_CL_USC_result_for_CV_system}
In the case of a CV system, if we consider the free Hamiltonian $\hat{H}_S$ to be given by
\begin{align}
\hat{H}_S = \frac{\hat{p}^2}{2m} + V(\hat{q}),
\end{align}
with $\hat{A} \equiv \hat{q}$, then the projectors $\hat{P}_i$ in Eq.~\ref{eqn_CL_USC_reduced_H} will be replaced by $\ket{q}\bra{q}$ and the summation over $i$ will be replaced by an integral over the eigenvalue $q$, giving rise to
\begin{align}
\Tilde{H} &= \int dq \braket{q | \hat{H}_S | q} \ket{q}\bra{q} \label{eqn_USC_H_CV_system}\\
&= \int \frac{1}{2m}\braket{q | \hat{p}^2 | q} \ket{q}\bra{q} dq + \int V(q) \ket{q}\bra{q} dq. \label{eqn_kinetic_n_potential_operators}
\end{align}
Since both the terms in Eq.~\ref{eqn_kinetic_n_potential_operators} commute with each other, the corresponding USC-MFGS (Eq.~\ref{eqn:general USC state}) is given as
\begin{align}
\hat{\rho}_{\text{USC}} &= Z^{-1} \exp \left\{ -\beta \Tilde{H} \right\}\\
&= Z^{-1} e^{-\beta V(\hat{q})} \exp \left\{ - \frac{\beta}{2m} \int \braket{q | \hat{p}^2 | q} \ket{q}\bra{q} dq \right\}.
\end{align}
Now, we have $\braket{q | \hat{p}^2 | q} = \braket{q + a| \hat{p}^2 | q + a}$, for all real $a$. Therefore, $\exp \left\{ - \frac{\beta}{2m} \int \braket{q | \hat{p}^2 | q} \ket{q}\bra{q} dq \right\}$ is proportional to the identity operator, and hence can be absorbed into the trace. We therefore get the USC-MFGS for CV system as \cite{ankerhold2003phase}
\begin{align}
    \hat{\rho}_{\text{USC}} &= Z'^{-1} e^{-\beta V(\hat{q})}.
\end{align}

\end{appendix}
%\nocite{*}
\bibliography{./citation}% Produces the bibliography via BibTeX.

\begin{thebibliography}{67}
\bibitem [{aba()}]{abandtilda}%
  \BibitemOpen
  \href@noop {} {}\bibinfo {note} {Since $\hat{A}$ and $\hat{\phi}$ commute, let, in the above equation, $\ket{a,a'}$ and $\ket{b,b'}$ be the corresponding eigenvectors of $\hat{\phi}$ and $\hat{A}$ with eigenvalues $a, b$ and $a', b'$, respectively. Then, if $a' \neq \tilde{a}$ or $b' \neq \tilde{a}$, then, as remarked in \autoref{sec:USC-MFGS in CL model}, for the corresponding action we have, $S_S[\tilde{a}, \phi(t)] = \infty$, effectively making $\hat{\rho}_{\text{USC}}$ diagonal in the basis of $\hat{A}$.}\BibitemShut {Stop}%
\bibitem [{del()}]{deltaexplaination}%
  \BibitemOpen
  \href@noop {} {}\bibinfo {note} {$\Delta$ quantifies, for the potential $V_k(x, y)$, for a fixed value of $y$, a (possibly) arbitrarily large but finite length scale about the global minima point $x = \Tilde{x}(y)$ such that for $|x - \Tilde{x}(y)| > \Delta$, the value of the potential $V_k(x, y)$ is so large that it does not influence the MFGS path integral.}\BibitemShut {Stop}%
\end{thebibliography}

\end{document}